\documentclass[12pt]{article}
\newlength{\vshift}
\input epsf.tex
\newlength{\hshift}

\usepackage{amssymb}
\usepackage{amsmath}
\usepackage{amsthm}
\usepackage{amsopn}

\def\beq{\begin{equation}}
\def\eeq{\end{equation}}
\def\bea{\begin{eqnarray}}
\def\eea{\end{eqnarray}}

\def\h{\hat}
 \def\L{\Lambda}
\begin{document}

 \vspace*{3cm}

\begin{center}

{\huge Effective  Field Theory of a Locally Noncommutative
Space-Time and Extra Dimensions}

\vskip 4em

{ {\bf B.~Mirza} \footnote{e-mail: b.mirza@cc.iut.ac.ir  }\: and \:
{\bf M.~Zarei} \footnote{e-mail: zarei@ph.iut.ac.ir }}

\vskip 1em

Department of Physics, Isfahan University of Technology, Isfahan
84156-83111, Iran
 \end{center}

 \vspace*{1.5cm}

\begin{abstract}
We assume that the noncommutativity starts to be visible
continuously from a scale $\Lambda_{NC}$. According to this
assumption, a two-loop effective action is derived for
noncommutative $\phi^{4}$ and $\phi^{3}$ theories from a
Wilsonian point of view. We show that these effective theories
are free of UV/IR mixing phenomena. We also investigate the
positivity constraint on coefficients of higher dimension
operators present in the effective theory. This constraint makes
the low energy theory to be UV completion of a full theory.
Finally, we discuss
 noncommutativity and extra dimensions. In our effective
theories formulated on noncommutative extra dimensions, if the
campactification scale $\Lambda_{c}$ is less than the scale
$\Lambda_{NC}$, the theory will not suffer from UV/IR mixing.
\end{abstract}
%%%%%%%%%%%%%%%%%%%%%%%%%%%%%%%%%%%%%%%%%%%%%%%%%%%%%%%%%%%%%%%%%%%%%%%%%%%%%%%%%%%%%%%%%%%%%%%%%%%%%%%%%%%%%%%%%%%%%%%%%%%
\newpage
\section{Introduction}
\label{A} Models of noncommutative space-time have become
increasingly popular recently and are believed to be reasonable
candidates for Planck scale physics. The main idea is that in the
scales where quantum theory and general relativity are no longer
independent, the notion of a point in space-time becomes meaningless
and a finite minimal length or uncertainty relation
 will have to be postulated for the coordinate functions in order to
prohibit the localization of points with arbitrary high precision.
Models with uncertainty relations are usually implemented by
considering noncommutative coordinate operators:
$[x^{\mu},x^{\nu}]=i\theta^{\mu\nu}$ where $\theta^{\mu\nu}$ is a
real and antisymmetric tensor and is of dimension $(mass)^{-2}$.
Field theories on noncommutative space-time have received a great
deal of attention, not least because they arise naturally in a
particular Seiberg-Witten limit \cite{1} of string theory (see
\cite{2,3}) for reviews. Noncommutative field theories are
constructed from conventional (commutative) field theory by
replacing in the Lagrangian the usual multiplication of fields with
the $\star$-product of fields,
 \beq
(\varphi\star\varphi)(x)=\varphi(x)e^{\frac{i}{2}\theta^{\mu\nu}\overleftarrow{\partial}_{\mu}\overrightarrow{\partial}_{\nu}}\varphi(x)
\eeq

 The parameter $\theta^{\mu\nu}$ then appears in the vertices of
 perturbation theory and defines a second mass scale in the
 theory, called noncommutativity scale $\Lambda_{NC}$. In general, for
 any noncommutative field theory, the loop diagrams can be
 classified into the so-called ``planar'' and ``non-planar'' graphs
 where the planar part of diagrams show the same UV divergence
 structure as the corresponding commutative theory, while the
 non-planar pieces are UV finite. It is natural to ask whether
 these theories are renormalizable. The UV renormalizability of
 noncommutative field theories have been investigated using
 counter term \cite{4,a5,a6,a7} and using Wilson-Polchinski method
 \cite{5,a8,a9}. On the other hand, perturbation calculations show the noncommutative field
 theories to be afflicted from an endemic nontrivial mixing of
 ultraviolet (UV) and infrared (IR) divergences \cite{6}. As a
 consequence, the Wilsonian approach to field theory seems to break
 down. Here the high energy modes of the noncommutative theory can
 not be decoupled because integrating out high energy degrees of
 freedom produces unexpected low energy divergences induced in
 the infrared operators of negative dimension \cite{6,7}. Generally
 speaking,
  in every quantum field theory formulated on a
 noncommutative space-time, we will have problems defining low
 energy Wilsonian effective action since UV and IR scales do
 not decouple. This lack of decoupling of different scales in
 the theory might pose serious problems for phenomenology, since
 noncommutative effects can show up at low energies interfering
 with standard model predictions \cite{7,8,9}. However, the ordinary soft and collinear
 divergences are decoupled from noncommutative IR divergences and there is no mixing between them
 \cite{10}. Recently  the renormalizability of noncommutative
 field theories have been investigated by introducing a harmonic oscillatory term in the free
 part of Lagrangian. \cite{11,12,a10,a11}. Here in this work, we propose a new
 approach to recover the Wilsonian effective action. Our main
 assumption is that there is an infrared cut-off $\Lambda_{NC}$
 under which noncommutativity can not be probed and space-time becomes commutative continuously. At energies below
 $\Lambda_{NC}$, physics of all degrees of freedom is governed by a
 commutative theory. The scale $\Lambda_{NC}$ can be equal
 to the scale of validity of an ordinary quantum field theory
 description of nature that is probably  around $(1-100)TeV$
 \cite{13}. So we can probe a truly noncommutative behavior only
 in the range $\Lambda_{NC}<k<\Lambda_{0}$. Here, the ultraviolet cut-off
$\Lambda_{0}$ is the scale beyond which the gravitational effects
are comparable to those of the rest of the fundamental interactions,
i.e. $\Lambda_{0}\sim M_{P}$(Planck mass). Beyond the scale
$\Lambda_{0}$, the noncommutative theory breaks down and physics is
sensitive to the UV completion of the theory. Now the Wilsonian
effective action can be obtained for noncommutative $\phi^{4}$ and
$\phi^{3}$ by integrating out the noncommutative high energy fields.
The feature of these effective theories is that they are free from
noncommutative IR divergences or UV/IR mixing. These results can be
extended to all orders of $g$. In the last section, we will discuss
the noncommutative extra dimension by employing the $\phi^{3}$
example on $R^{1,3}\times T_{\theta}^{2}$. The paper is organized as
follows: In section 2, we will first review the renormalizability of
noncommutative $\phi^{4}$ theory using Wilson-Polchinski method.
Section 3 is devoted to a detailed analysis of deriving Wilsonian
effective action for noncommutative $\phi^{4}$ and $\phi^{3}$
theories. We also discus the positivity constraint \cite{14} on
coefficients of higher dimensional terms. In section 4, we discuss
the noncommutative extra dimensions and how the UV/IR mixing can be
canceled out in the perturbation calculation.
%%%%%%%%%%%%%%%%%%%%%%%%%%%%%%%%%%%%%%%%%%%%%%%%%%%%%%%%%%%%%%%%%%%%%%%%%%%%%%%%%%%%%%%%%%%%%%%%%%%%%%%%%%%%%%%%%%%%%%%%%%%
\section{Renormalizability of the noncommutative $\phi^{4}$ theory}

In this section we first review the renormalization of
noncommutative $\phi^{4}$ theory \cite{5,a8,a9} using the Wilsonian flow
equation formulated by Polchinski \cite{15,16}. Then we will have a
short review of  Grosse and Wulkenhaar method \cite{11,12,a10,a11}. At
first
 our starting point is the path integral
\bea Z[J]=e^{W[J]}=\int {\cal D} \phi
\!\!\!\!\!\!\!\!&&\exp\left\{-\frac{1}{2} \int\frac{d^4p}{(2\pi)^4}
\phi(p) D^{-1} \phi(-p) - S_{\mathrm int}[\phi] \right.
\nonumber\\&&
 \:\:\:\:\: +\left. \int\frac{d^4p}{(2\pi)^4}
\phi(p)J(-p)\right\}\ \eea where
\bea && S_{\mathrm int}=\frac{1}{4!}\:
g  \int\frac{d^4p}{(2\pi)^4} \phi\star\phi\star\phi\star\phi,\nonumber\\ && D(p)=(p^{2}+m^{2})^{-1} \eea
 Due to
star product of fields, the Feynman rules for the interaction vertex
changes to \bea
\Gamma^{(4)}(p_1,p_2,p_3,p_4)\!\!\!\!\!\!\!\!&&= g^2 h(p_1,p_2,p_3,p_4) \nonumber\\
&& \equiv \frac{g^2}{3}\big[\cos(\frac{1}{2}p_{1}\theta
p_{2})\cos(\frac{1}{2}p_{3}\theta p_{4}) +
\cos(\frac{1}{2}p_{1}\theta p_{3})\cos(\frac{1}{2}p_{2}\theta
p_{4})\nonumber\\ &&
\:\:\:\:\:\:\:\:\:\:\:\:\:\:\:\:\:\:\:\:\:\:\:\:+
 \cos(\frac{1}{2}p_{1}\theta
p_{4})\cos(\frac{1}{2}p_{2}\theta p_{3}) \big] \eea
 Working
with this vertex leads to some problems. For instance, consider the
one loop mass correction
 \bea  \Sigma_{\Lambda_0}(p)\!\!\!\!\!\!\!&&= -
\frac{g^2}{6}\int\frac{d^4q}{(2\pi)^4} \Theta(\Lambda_0^2-q^2)
\frac{1}{q^2+m^2}\big[\: 2+\cos( q\cdot\tilde{p})\:\big]
\nonumber\\&& =-\frac{g^2}{32 \pi^2}\big[\:\frac{2}{3}\Lambda_0^2 +
\frac{4}{3~ \tilde{p}^2 }
\big(1-J_0(\Lambda_0~\tilde{p})\big)\big]\;,\;\;\;(m^2 \ll
\mathrm{min}\{\Lambda_0^2,1/\tilde{p}^2\}) \eea

\noindent where $\tilde{p}^{\mu}\equiv\theta^{\mu\nu}p^\nu$. In
equation (5), the second term is contributed by the non-planar
graphs containing an extra phase factor and a new kind of IR
divergence induced by the effective UV cut-off (UV/IR mixing)
arising from this term \cite{6}. To get the RG equation, we first
introduce  an UV and IR cut-offs $\Lambda_{0}$ and $\Lambda$ into
the theory by making the following substitution \beq D(p)\rightarrow
D_{\Lambda, \Lambda_0}(p) \equiv D(p) K_{\Lambda, \Lambda_0}(p) \eeq
 \noindent where
$K_{\Lambda, \Lambda_0}(p)$ is equal to one in the region
$\Lambda^{2}<p^{2}<\Lambda^{2}_{0}$ and vanishes rapidly outside.
The substitution above defines $Z_{\Lambda, \Lambda_0}$ and
$W_{\Lambda, \Lambda_0}$, the generating functionals of Green
functions in which only momenta between $\Lambda$ and $\Lambda_0$
have been integrated out. By means of a simple recipe given in
\cite{5,a8,a9}, we obtain the RG equation for a given Green function.
These equations form an infinite system of coupled ordinary
differential equations. As an example, the evolution equation for
self-energy is given by \beq  \Lambda\frac{\partial\;\;}{\partial
\Lambda}\Sigma_{\Lambda, \Lambda_0}(p) = \frac{1}{2} \int\frac{d^4
q}{(2\pi)^4}\frac{S_{\Lambda, \Lambda_0}(q)}{q^2+m^2}
\Gamma^4_{\Lambda, \Lambda_0}(q,p,-p,-q)\ \eeq
 Where
 \bea
\frac{S_{\Lambda, \Lambda_0}(q)}{q^2+m^2} && \equiv
\Lambda\frac{\partial\;\;} {\partial \Lambda}
\left.\frac{1}{(q^2+m^2)
K_{\Lambda,\Lambda_0}(q)^{-1}+\Sigma_{\Lambda^\prime,\Lambda_0}(q)}
\right|_{\Lambda^\prime=\Lambda}\,,\nonumber\\
&& =\frac{1}{q^2+m^2} \frac{1}{\left[ 1+\frac{\Sigma_{\Lambda,
\Lambda_0}(q)}{q^2+m^2} K_{\Lambda,\Lambda_0}(q)  \right]^2}
\L\frac{\partial\;\;} {\partial \Lambda}K_{\Lambda,\Lambda_0}(q)\
\eea

In order to study the UV renormalization, the relevant couplings
i.e. those with non-negative mass dimensions, have to be isolated.
They are
 \beq \gamma_2(\L)\equiv \left.\frac{d
\Sigma_{\L, \L_0}(p)}{d p^2}\right|_{p^2=p_0^2},\;\;\;\;
\gamma_3(\L)\equiv \left.\Sigma_{\L,
\L_0}(p)\right|_{p^2=p_0^2},\;\;\;\; \gamma_4(\L)\equiv
\frac{\Gamma^{4}_{\L, \L_0}(\bar{p}_1,\ldots,
\bar{p}_{4})}{h(\bar{p}_{1},...,\bar{p}_{4})} \eeq

\noindent where the momentum $\bar{p}_{i}$ is chosen such that
$\bar{p}_{i}\cdot\bar{p}_{j}=p_{0}^{2}(\delta_{ij}-\frac{1}{4})$ in
which $p_{0}^{2}$ is the renormalization scale. The two- and
four-point function can be rewritten as \bea  && \Sigma_{\L,
\L_0}(p) = \gamma_3(\L) + (p^2-p_0^2) \gamma_2(\L)+
\Delta^2_{\L, \L_0}(p) \nonumber \\
&&  \Gamma^{4}_{\L, \L_0}(p_1,\ldots, p_{4})=\gamma_4(\L)
+\Delta^4_{\L, \L_0}(p_1,\ldots, p_{4})\eea where $\Delta^2_{\L,
\L_0}$ and $\Delta^4_{\L, \L_0}$ are irrelevant operators with all
$\Gamma^{2n}_{\L, \L_0}$'s with $n\geq 2$. Integrating RG equations
with respect to cut-off parameter from $0$ to $\Lambda$ for the
relevant operators and from $\Lambda$ to $\Lambda_{0}$ for
irrelevant operators leads to a set of coupled integral equations
with the boundary conditions \beq
\gamma_2(0)=0,\,\,\,\gamma_3(0)=p^{2}_{0},\,\,\,\gamma_4(0)=g^2
\,\,\,\, (at \,\, physical \,\, point \,\, \Lambda=0) \eeq and \beq
\Delta^2_{\L_0, \L_0}(p)=\Delta^4_{\L_0, \L_0}(p_1,\ldots,
p_{4})=\Gamma^{2n}_{\L_0,
\L_0}(p_1,\ldots,p_{2n})=0\;\;\;\;\;\;\;\;\;\;\; (n > 2)\; \eeq

The above boundary conditions indicate the renormalization
conditions. For instance, at one loop, the relevant and irrelevant
couplings of the two-point functions contribute as in the following:
\beq \gamma_3(\Lambda) = -\frac{g^2}{32 \pi^2}\left[\frac{2}{3} \L^2
+ \frac{4}{3 \tilde{p}_0^2}(1-J_0(\Lambda \tilde{p_0}))\right], \eeq
\beq \gamma_2(\Lambda) = \frac{g^2}{24
\pi^2}\frac{1}{p_0^2\tilde{p_0}^2}
 \left[1 - J_0(\L \tilde{p_0})-\frac{\Lambda \tilde{p_0}}{2}
J_1(\Lambda \tilde{p_0})\right],\;\;   \eeq \bea \label{D1L}
&&\Delta^2_{\L,\L_0}=\frac{g^2}{24 \pi^2} \left\{\frac{J_0(\L
\tilde{p})- J_0(\L_0 \tilde{p})}{\tilde{p}^2}- \frac{J_0(\L
\tilde{p_0})- J_0(\L_0 \tilde{p_0})}{\tilde{p_0}^2}
\right.\nonumber\\
&&\;\;\;\;\;\;\;\;\;\;\;\;\;\left.
\frac{p^2-p_0^2}{p_0^2\tilde{p_0}^2}\left[ J_0(\L
\tilde{p_0})-J_0(\L_0 \tilde{p_0}) +\frac{\L \tilde{p_0}}{2}J_1(\L
\tilde{p_0})- \frac{\L_0 \tilde{p_0}}{2}J_1(\L_0
\tilde{p_0})\right]\right\}.\eea

The proof of renormalizability is based on the scaling behavior of
relevant and irrelevant operators. In order to simplify the power
counting, we assume the following relations to hold between scales,
\bea \tilde{p}\Lambda_{0}\gg 1 \:\:\: , \:\:\:
\Lambda_{NC}\ll\Lambda\ll\Lambda_{0} \eea The UV renormalizability
can be proved using induction at any order of perturbation. At
one-loop level, the UV cut-off $\Lambda_{0}$ is removed from
operators \cite{5,a8,a9}. Now at $l$-loop, we assume the
$\Lambda_{0}$-independent scaling using power law behavior for
operators; for example, \beq \label{SCALINGS}
\gamma_2^{(l)}\sim\gamma_4^{(l)}\sim\left|\Delta^4_{(l)}\right|_\Lambda
=O(1) \;,\;\;\;\;\gamma_3^{(l)}\sim \left|\Delta^2_{(l)}\right|_\L
=O(\L^2)\; \eeq
Using the renormalization group equation, it can be
shown that at $(l+1)$-loop, the relevant and irrelevant couplings
are finite in the $\Lambda_0\rightarrow {\infty}$ (UV) limit. So the
perturbation renormalizability at any order in perturbation theory
is proved. On the other hand, when the IR region
($\tilde{p}\Lambda_{0}\ll 1$) comes under scrutiny, the Wilsonian
picture breaks down because of the UV/IR phenomena, and the low
energy predictions under the scale
$\frac{\Lambda_{NC}}{\Lambda_{0}}$ become highly sensitive to the
details of the ultraviolet sector of the theory. In order to write
down a Wilsonian effective action which correctly describes the low
momentum behavior of the theory, one may add a new degree of freedom
$\chi$ to the commutative action \cite{6}. Then, the modified
Wilsonian action gets the following form \beq S_{\mathrm
eff}'(\Lambda)=S_{\mathrm eff}(\Lambda) +\int
d^{4}x\Big(\frac{1}{2}\:(\tilde{\partial}\chi)^{2}+\frac{1}{2}\Lambda^{2}(\tilde{\partial}^{\:2}\chi)^{2}
+i\frac{1}{\sqrt{96\pi}}\:g\chi\phi
  \Big)\eeq
Integrating out the new mode in the region $\tilde{p}\Lambda_{0}\ll
1$ gives the quadratic infrared singularity
$\frac{1}{\tilde{p}^{2}}$ appearing in the self energy diagrams (5)
\cite{5,a8,a9,6}.

There is also a new approach developed by Grosse and Wulkenhaar in
\cite{11,12,a10,a11} where it is shown that noncommutative four dimensional
field theories are renormalizable to all orders of perturbation.
This method also is a solution to UV/IR mixing problem. This work is
performed by adding a new harmonic oscillator term to the free part
of the action,

 \beq \nonumber
S(\phi)=\int d^{4}x \big(\:
\partial_{\mu}\phi\star\partial^{\mu}\phi+\frac{\Omega^{2}}{2}(\tilde{x}_{\mu}\phi)\star(\tilde{x}^{\mu}\phi)
+\frac{m^{2}}{2}\phi\star\phi \:\big)
 \eeq
where $\tilde{x}_{\mu}=2(\theta^{-1})_{\mu\nu}x^{\nu}$ and
$\Omega$ is a dimensionless constant. This additional new term
causes that the theory to be invariant under a sort of duality
transformation between positions and momenta at $\Omega=1$ (away
from this special point, it is more precise to say that the model
is covariant) \cite{17}. A main difference between this theory
and ordinary noncommutative theories is that a free particle is
affected by noncommutativity of space-time i.e. its propagator
and also dispersion relation are modified without any
interaction. In momentum space, working with this theory and
specially deriving the propagators is quite hard and is done only
in a matrix representation form. Although if one work in a
x-space formalism \cite{18}, it seems that to some extent, this
cumbersome theory is simplified.

 In the following section, we explain
a new idea which is a physical realization of mathematical view
point of reference \cite{19} and derive an effective action.
However, in this paper we do not use a Minkowski space-time and
perform our calculation with a Euclidean signature. In this
method we introduce a scale where space starts to be
noncommutative, and by integrating out the noncommutative fields
with momenta beyond this scale, we get an effective theory which
is free from UV/IR mixing. The harmonic oscillator term
$\frac{\Omega^{2}}{2}(\tilde{x}_{\mu}\phi)\star(\tilde{x}^{\mu}\phi)$
does not change our result because the  integration domain begins
from a scale where beyond it this term become irrelevant. So we
only use the usual noncommutative field theory as we only need to
integrate out high energy modes.

%%%%%%%%%%%%%%%%%%%%%%%%%%%%%%%%%%%%%%%%%%%%%%%%%%%%%%%%%%%%%%%%%%%%%%%%%

\section{Noncommutative Wilsonian effective action}
\subsection{$\phi^{4}$ theory}

In ordinary field theory, low-energy dynamics (large distances)
does not depend on details of high energy dynamics (short
distance). Using Wilson's method, one can decouple the high
energy modes and obtain an effective field theory which has the
same infrared behaviour as the underlying fundamental theory. In
noncommutative field theory, the high energy modes are not
decoupled  because of UV/IR phenomena and the Wilsonian approach
to effective field theory seems to break down. However, making a
new assumption we start to rederive a well-behaved Wilsonian
effective action for noncommutative $\phi^{4}$ and $\phi^{3}$
theories. At first, we turn to noncommutative $\phi^{4}$ and
write the generating functional for it in the $d$-dimensional
Euclidean space. Here we consider the Grosse-Wulkenhaar term,
 \beq Z=\int {\cal
D}\phi\exp \Big(-\int d^{d}x
\Big[\frac{1}{2}\partial_{\mu}\phi\partial^{\mu}\phi+m^{2}\phi^{2}+
\frac{\Omega^{2}}{2}(\tilde{x}_{\mu}\phi)\star(\tilde{x}^{\mu}\phi)
  +\frac{1}{4!}g\phi\star\phi\star\phi\star\phi\Big]\Big)
 \eeq

We now divide the integration variables $\phi(k)$ into two groups,
those with $|k|<\Lambda=\Lambda_{NC}$ and those with
$\Lambda=\Lambda_{NC}\leq|k|<\Lambda_{0}$. The main assumption in
our calculation is to suppose that the noncommutative behaviour
can be detected in the range of energies beyond $\Lambda_{NC}$
and that below this scale the theory becomes to be continuously
commutative . So we define the fields $\phi$ with the momentum
$|k|<\Lambda_{NC}$ to have an ordinary (commutative) behavior and
the fields $\h{\phi}$ with the momentum
$\Lambda_{NC}\leq|k|<\Lambda_{0}$ to have noncommutative
behavior. Then we can replace the old field $\phi$ by
$\phi+\h{\phi}$.

 The Eq.(19) without the harmonic oscillator term is given by,
\beq Z=\int {\cal D}\phi\int{\cal D}\h{\phi}\exp \Big(-\int d^{d}x
\Big[\frac{1}{2}(\partial_{\mu}\phi+\partial_{\mu}\h{\phi})^{2}+\frac{1}{2}m^{2}(\phi+\hat{\phi})^{2}
  +\frac{1}{4!}g(\phi+\h{\phi})_{\star}^{4}\Big]\Big) \eeq

Here by expanding the terms in (20), a new problem will appear.
 Because of previous assumptions about $\phi$ and $\hat{\phi}$
  we can write \bea &&\phi\star\phi=\phi\phi \nonumber\\
&& \phi\star\hat{\phi}=\phi\hat{\phi} \nonumber\\ &&
\hat{\phi}\star\hat{\phi}=\hat{\phi}\star\hat{\phi}  \eea i.e. the
product becomes commutative one whenever one of fields which is
being multiplied has a momentum which is less than the
noncommutative scale. The ambiguity is appeared when for example
we consider a product such as
$(\phi+\hat{\phi})\star(\phi+\hat{\phi})\star(\phi+\hat{\phi})$.
Since

\beq(\phi+\hat{\phi})\star[(\phi+\hat{\phi})\star(\phi+\hat{\phi})]=
[(\phi+\hat{\phi})\star(\phi+\hat{\phi})]\star(\phi+\hat{\phi})\eeq
it does not matter which one we take to expand. If we expand the
left side so the terms that are linear in $\hat{\phi}$ are \bea &&
\phi\star(\phi\star\hat{\phi}) \nonumber\\ && \phi\star(\hat{\phi}\star\phi) \nonumber\\
&& \hat{\phi}\star(\phi\star\phi) \eea The first two can be
evaluated easily according to the rule above and give
$\phi^{2}\hat{\phi}$. However the last one gives \beq
\hat{\phi}\star(\phi\star\phi)=\hat{\phi}\star(\phi^{2})  \eeq Now
the modes in $\phi^{2}$ will have momentum which is a sum of those
coming from the two $\phi$'s. So even if $\phi$ contain only modes
with low momentum, the $\phi^{2}$ will contain modes with momentum
larger than the scale $\Lambda_{NC}$ and so one can not assume
$\phi^{2}$ has commutative behavior.

In order to solve this ambiguity it seems that in a more
fundamental treatment we must find a symmetry breaking mechanism
to deal with the high and low energy modes. However, in lack of
such a theory we may continue with some approximations in our
calculations. For example, in the terms such as $\phi^{2}$ we can
approximately ignore some modes whose momentum become larger than
$\Lambda_{NC}$ in comparison with high momentum modes of
$\hat{\phi}$ and then we have,

\bea Z=\int {\cal D}\phi
e^{-\int\cal{L}(\phi)}&&\!\!\!\!\!\!\!\!\!\!\!\!\!\int{\cal
D}\h{\phi}\exp \Big(-\int d^{d}x
 \Big[\frac{1}{2}(\partial_{\mu}\h{\phi})^{2}+  \frac{1}{2} m^{2}\hat{\phi}^{2}+
 \frac{\Omega^{2}}{2}(\tilde{x}_{\mu}\hat{\phi})\star(\tilde{x}^{\mu}\hat{\phi})\nonumber\\
&&
 \!\!\!\!\!\!\!\!\!\!\!\!\!\!\!\!\!\!\!\!\ +g\:(\frac{1}{6}\phi^{3}\hat{\phi}+\frac{1}{4}\phi^{2}\hat{\phi}\star\hat{\phi}
   +\frac{1}{6}\phi\:\hat{\phi}\star\hat{\phi}\star\hat{\phi}
   +\frac{1}{4!}\hat{\phi}\star\hat{\phi}\star\hat{\phi}\star\hat{\phi})\Big]\Big)
   \eea
\noindent where $\cal{L}(\phi)$ is the commutative Lagrangian. Now
we can ignore the
$\frac{\Omega^{2}}{2}(\tilde{x}_{\mu}\hat{\phi})\star(\tilde{x}^{\mu}\hat{\phi})$
term since this term is highly irrelevant in short distances or
equivalently in energies above $\Lambda_{NC}$.

\noindent Integrating out all $\h{\phi}$s with the momentum
$\Lambda_{NC}\leq|k|<\Lambda_{0}$, transforms (25) into the
following form \beq Z=\int{\cal D}\phi
e^{-S_{eff}(\phi;\Lambda_{NC})} \eeq \noindent  where
$S_{eff}(\phi;\Lambda_{NC})$ is the deformed Wilsonian effective
action. The corresponding Lagrangian density will finally  be
rewritten as follows:
 \bea {\cal
L}_{eff}(\phi;\Lambda_{NC})\!\!\!\!\!\!\!\!&&=\frac{1}{2}Z(\Lambda_{NC})\partial_{\mu}\phi\partial^{\mu}\phi
+\frac{1}{2}m^{2}(\Lambda_{NC})\phi^{2}+\frac{1}{4!}g(\Lambda_{NC})\phi^{4}\nonumber\\
&&
\:\:\:\:\:\:\:\:\:\:\:\:\:\:\:\:\:\:\:\:\:\:+\sum_{i}C_{n,d,i}(\Lambda_{NC}){\cal
O}_{n,d,i} \eea \noindent where the terms ${\cal O}_{n,d,i}$ are all
irrelevant local operators with the dimension $D=2n+d\geq \:6$ and
consist of all the terms that can be constructed out of an even
number $2n$ of $\phi$ fields with $d$ number of derivatives acting
on them. The index $i$ keeps track of operators with the same values
of $n$ and $d$ that are not equivalent after integration  by parts.
We will show explicitly the dependence of coefficients $Z\:, m^{2},
g$ and $C_{n,d,i}$'s on the noncommutative parameter $\theta$. The
crucial feature of (27) is that the UV/IR mixing problem is solved
naturally in it. Also the coefficients $C_{n,d,i}$ satisfy the
positivity constraint discussed in \cite{14} in order for the low
energy effective theory to be UV complete into a full theory. Now
before turning to calculation parameters, note that we assume
$m^{2}\ll \Lambda_{NC}^{2}$ and treat the mass term
$\frac{1}{2}m^{2}\hat{\phi}$ as a perturbation. To perform the
perturbation calculation, we expand the exponential and use the
Wick's theorem. In a Feynman diagrams approach, the fields $\phi$
just contribute as an external line and the fields $\hat{\phi}$
contribute via internal lines and its propagator is defined as the
commutative ones, \beq
\big<\hat{\phi}(p)\hat{\phi}(k)\big>=\frac{1}{k^{2}}(2\pi)^{d}\delta^{d}(k+p)\Theta(k)\eeq
\noindent where \beq \Theta(k)=\Big\{_{0\:\:\:\:\:\: otherwise}^{1
\:\:\:\:\:\: if \:\:\:\Lambda_{NC}\leq|k|<\Lambda_{0}} \eeq
 The diagrams that must be
evaluated for the case of the two external lines are drawn up to
order $g^{2}$ in Fig. 1. The thin line represents the external
fields $\phi$ and the dark line shows the internal fields
$\hat{\phi}$.

\begin{figure}\vspace{-5
cm }
\centerline{\epsfysize=6in\epsfxsize=8in\epsffile{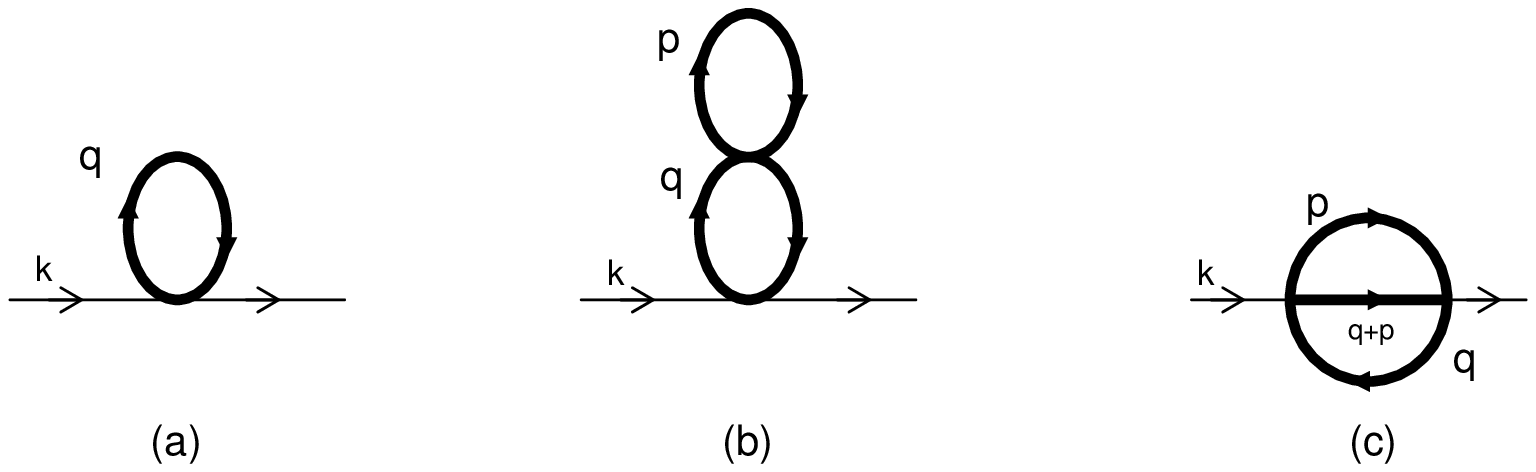}}\vspace{-5
cm }\caption{Two external line diagrams
 }\label{fig1}
\end{figure}

First, we calculate Fig. 1a resulting from expansion of
$\phi^{2}\hat{\phi}^{2}$ term in the exponent (25). We have
 \beq -\frac{1}{2}\int
d^{d}x\:\mu_{1}\:\phi^{2} \eeq Where \beq
\mu_{1}=\frac{g}{2}\int\frac{d^{d}q}{(2\pi)^{d}}\frac{1}{q^{2}}=
\frac{\pi^{\frac{d}{2}}g}{(2\pi)^{d}\Gamma(\frac{d}{2})}\:\frac{1}{d-2}\:\Big(\Lambda_{0}^{d-2}-\Lambda_{NC}^{d-2}\Big)
\eeq
 The second diagram (Fig. 1b) results from the term
$(\phi^{2}\hat{\phi}\star\hat{\phi})(\hat{\phi}\star\hat{\phi}\star\hat{\phi}\star\hat{\phi})$.
Then for this diagram we have \beq -\frac{1}{2}\int
d^{d}x\:\mu_{2}\:\phi^{2} \eeq Where
 \bea \mu_{2}\!\!\!\!\!\!\!\!&&= \frac{1}{2}\frac{g^{2}}{4!}\int \frac{d^{d}p}{(2\pi)^{d}}\frac{1}{p^{2}}
    \int \frac{d^{d}q}{(2\pi)^{d}}\frac{1}{(q^{2})^{2}}
 \nonumber\\
&& =\frac{1}{2}\frac{1}{4!}\big(
\frac{2\pi^{\frac{d}{2}}g}{(2\pi)^{d}\Gamma(\frac{d}{2})}\big)^{2}\frac{1}{d-2}\frac{1}{d-4}
\:\Big(\Lambda_{0}^{d-2}-\Lambda_{NC}^{d-2}\Big)\Big(\Lambda_{0}^{d-4}-\Lambda_{NC}^{d-4}\Big)
\eea

In the same way, for the remaining diagram (Fig. 1c) we will have

 \beq Fig.\:1c=-\frac{1}{2}\int d^{d}x\:\mu_{3}\:\phi^{2} \eeq

\noindent  where using the table of integrals \cite{20},
calculations will
 yield:

 \bea \mu_{3}\!\!\!\!\!\!\!\!\!&&=\big(\frac{g}{6}\big)^{2}\int\frac{d^{d}p}{(2\pi)^{d}}\frac{1}{p^{2}}\int \frac{d^{d}q}{(2\pi)^{d}}\frac{1+\cos p\:\theta q}{q^{2}(q+p)^{2}}
 \nonumber\\
&& =\big(\frac{g}{6}\big)^{2}\big[\:{\cal F}(1,1)+{\cal
G}(1,1;\theta)\:\big]
 \eea
The functions ${\cal F}(1,1)$ and ${\cal G}(1,1;\theta)$ can be
found in the appendix A. In the above equation, the factor
$\cos(p\theta k)$ is arising from Feynman rules derived from
noncommutative $\hat{\phi}^{3}$ vertex. Note that at this stage,
in Fig. 1c, we have considered the limit in which the external
momenta carried out by the factors $\phi$ are very small compared
to $\Lambda_{NC}$ and they can, therefore, be ignored. The
coefficients $\mu_{i}$ give corrections to the $m^{2}$ term in
$\mathcal{L}$ along the following lines:

 \bea
m^{2}(\Lambda_{NC})&&\!\!\!\!\!\!\!\!\!=m^{2}(\Lambda_{0})+\mu_{1}+\mu_{2}+\mu_{3}\nonumber\\
&& \!\!\!\!\!\!\!\!\!= m^{2}(\Lambda_{0})\nonumber\\
&& +
\:\:\frac{2\pi^{\frac{d}{2}}g(\Lambda_{0})}{(2\pi)^{d}\Gamma(\frac{d}{2})}\frac{1}{d-2}\Big(\Lambda_{0}^{d-2}-\Lambda_{NC}^{d-2}\Big)
\nonumber\\
&& +\:\: \frac{1}{2}\frac{1}{4!}\Big(
\frac{2\pi^{\frac{d}{2}}g(\Lambda_{0})}{(2\pi)^{d}\Gamma(\frac{d}{2})}\Big)^{2}\frac{1}{d-2}\frac{1}{d-4}
\:\Big(\Lambda_{0}^{d-2}-\Lambda_{NC}^{d-2}\Big)\Big(\Lambda_{0}^{d-4}-\Lambda_{NC}^{d-4}\Big)\nonumber\\&&
+ \:\:\frac{g^{2}(\Lambda_{0})}{36}\big[\:{\cal F}(1,1)+{\cal
G}(1,1;\theta)\:\big]
 \eea

\begin{figure}\vspace{-4
cm }
\centerline{\epsfysize=5in\epsfxsize=7in\epsffile{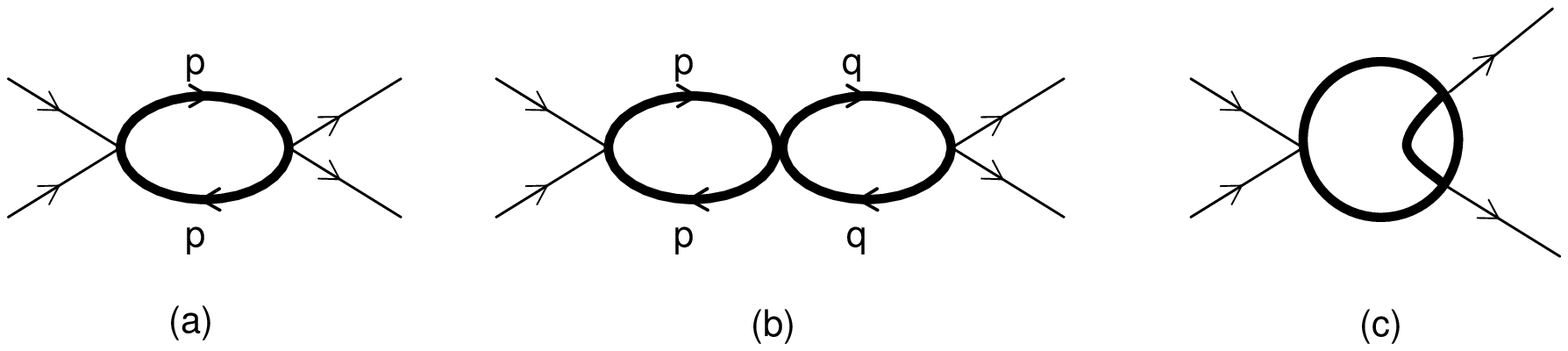}}\vspace{-6
cm }\caption{Four external line diagrams
 }\label{fig2}
\end{figure}

Let us now evaluate the diagrams with four external lines, up to
order $g^{3}$ as drawn in Fig. 2.
 The first loop diagram
is contributed as follows

 \beq Fig.\: 2a=-\frac{1}{4!}\int d^{d}x
\:\zeta_{1}\:\phi^{4} \eeq

\noindent where

\bea
\zeta_{1}=4!\big(\frac{g}{4}\big)^{2}\int\frac{d^{d}p}{(2\pi)^{d}}\frac{1}{(p^{2})^{2}}&&\!\!\!\!\!\!\!=
\frac{3\pi^{\frac{d}{2}}g^{2}}{(2\pi)^{d}\Gamma(\frac{d}{2})}\frac{1}{d-4}\big(\Lambda_{0}^{d-4}-\Lambda_{NC}^{d-4}\big)\nonumber\\&&
\!\!\!\!\!\!=
\:\:\:\frac{3g^{2}}{16\pi^{2}}\log\frac{\Lambda_{0}}{\Lambda_{NC}}
\:\:\:\:\:\:\:\:\:(d\rightarrow 4)\eea

\noindent Similarly, for the remaining two loop diagrams, we have

\beq Fig.\: 2b=- \frac{1}{4!}\int d^{d}x \:\zeta_{2}\:\phi^{4} \eeq

\noindent where
 \bea
\zeta_{2}&&\!\!\!\!\!\!\!\!\!\!=\frac{g^{3}}{16}\:\int\frac{d^{d}p}{(2\pi)^{d}}\frac{1}{(p^{2})^{2}}
\int\frac{d^{d}p}{(2\pi)^{d}}\frac{1}{(p^{2})^{2}}\nonumber\\&&
=\frac{g^{3}}{4}\Big[\:\frac{\pi^{\frac{d}{2}}g^{2}}{(2\pi)^{d}\Gamma(\frac{d}{2})}\frac{1}{d-4}\big(\Lambda_{0}^{d-4}-\Lambda_{NC}^{d-4}\big)\:\Big]^{2}
 \eea
and
 \beq Fig.\: 2c= -\frac{1}{4!}\int d^{d}x \:\zeta_{3}\:\phi^{4}
\eeq

\noindent with

 \bea \zeta_{3}\!\!\!\!\!\!\!\!\!&&=\frac{g^{3}}{12}\int\frac{d^{d}p}{(2\pi)^{d}}\frac{1}{(p^{2})^{2}}\int \frac{d^{d}q}{(2\pi)^{d}}\frac{1+\cos p\:\theta q}{q^{2}(q+p)^{2}}
 \nonumber\\
&& =\frac{g^{3}}{12}\big[\:{\cal F}(2,1)+{\cal G}(2,1;\theta)\:\big]
 \eea

\noindent The coefficients $\zeta_{1}$, $\zeta_{2}$ and $\zeta_{3}$
give a nonzero correction to $g$, \bea
g(\Lambda_{NC})&&\!\!\!\!\!\!\!\!=g(\Lambda_{0})+\zeta_{1}+\zeta_{2}+\zeta_{3}\nonumber\\&&
       \!\!\!\!\!\!\!\!= g(\Lambda_{0})+
\frac{3\pi^{\frac{d}{2}}g^{2}(\Lambda_{0})}{(2\pi)^{d}\Gamma(\frac{d}{2})}\frac{1}{d-4}\big(\Lambda_{0}^{d-4}-\Lambda_{NC}^{d-4}\big)\nonumber\\&&
+\:\frac{g^{3}}{4}\Big[\:\frac{\pi^{\frac{d}{2}}g^{2}}{(2\pi)^{d}\Gamma(\frac{d}{2})}\frac{1}{d-4}\big(\Lambda_{0}^{d-4}-\Lambda_{NC}^{d-4}\big)\:\Big]^{2}
     \nonumber\\&&  +\:\frac{g^{3}}{12}\big[\:{\cal F}(2,1)+{\cal
G}(2,1;\theta)\:\big]
           \eea

Dependence of the factors $m^{2}$ and $g$ to the noncommutative
parameter $\theta$ comes from ${\cal G}(n,m;\theta)$. This function
has a complicate form but for $d=4$, after performing the summation
on $r$ it is possible to see explicitly the soft behaviour of this
function in the limit $\theta\rightarrow 0$. So the loop calculation
is free of noncommutative IR divergences or UV/IR mixing which is
expected as a result of maintaining the assumption that
noncommutative effects appear to be visible from the scale
$\Lambda_{NC}$. Following this formalism, one can extend these
results to arbitrary orders of perturbation theory. We now turn to
terms with higher orders of $\phi$. For instance, the dominant
contribution to $C_{6,d,i}(\Lambda_{NC})$ is given by one and two
loop diagrams with six external lines, (See Fig. 3). Calculating
these diagrams, we find the coefficients $C_{6,d,i}(\Lambda)$ with a
negative mass dimension. In this part, the diagrams in Fig.\:3 are
calculated,
\begin{figure}\vspace{-5
cm }
\centerline{\epsfysize=5in\epsfxsize=7in\epsffile{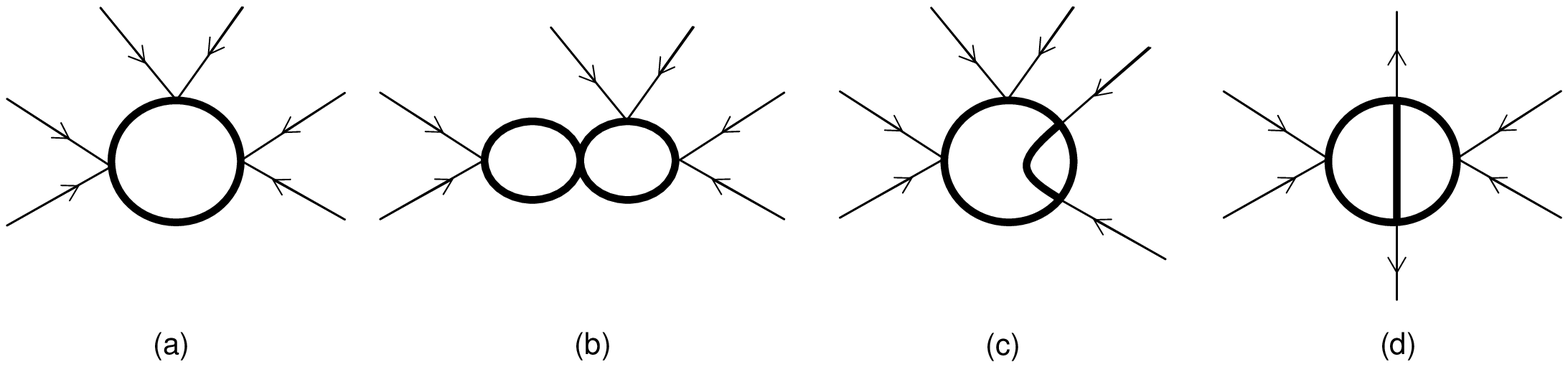}}\vspace{-4
cm }\caption{Six external line diagrams
 }\label{fig3}
\end{figure}

 \beq Fig.\:
3a=-\frac{1}{6!}\int d^{d}x \lambda_{1}\phi^{6} \eeq
 \beq
Fig.\: 3b=-\frac{1}{6!}\int d^{d}x \lambda_{2}\phi^{6} \eeq \beq
Fig.\: 3c=-\frac{1}{6!}\int d^{d}x \lambda_{3}\phi^{6} \eeq \beq
Fig.\: 3d=-\frac{1}{6!}\int d^{d}x \lambda_{4}\phi^{6} \eeq
 Where
$\lambda_{1}$,...,$ \lambda_{4}$ are defined as follows \bea
\lambda_{1}\equiv
C_{6,0,1}&&\!\!\!\!\!\!\!\!=6!(\frac{g}{4})^{3}\int\frac{d^{d}q}{(2\pi)^{d}}(\frac{1}{q^{2}})^{3}\nonumber\\&&
\!\!\!\!\!\!\!\!
=6!(\frac{g}{4})^{3}\frac{2\pi^{\frac{d}{2}}}{(2\pi)^{d}\Gamma(\frac{d}{2})}\:\frac{1}{d-6}\:(\Lambda_{0}^{d-6}-\Lambda_{NC}^{d-6})
\eea
 \bea \lambda_{2}\equiv
C_{6,0,2}&&\!\!\!\!\!\!\!\!=2.6!\frac{g^{4}}{4!4^{3}}\int\frac{d^{d}p}{(2\pi)^{d}}\frac{1}{(p^{2})^{2}}\int\frac{d^{d}q}{(2\pi)^{d}}\frac{1}{(q^{2})^{3}}\nonumber\\&&
\!\!\!\!\!\!\!\!
=2.6!\frac{g^{4}}{4!4^{3}}\big(\frac{2\pi^{\frac{d}{2}}}{(2\pi)^{d}\Gamma(\frac{d}{2})}\big)^{2}\:\frac{1}{d-4}\frac{1}{d-6}\:(\Lambda_{0}^{d-4}-\Lambda_{NC}^{d-4})(\Lambda_{0}^{d-6}-\Lambda_{NC}^{d-6})
\eea \bea \lambda_{3}\equiv
C_{6,0,3}&&\!\!\!\!\!\!\!\!=\frac{5}{2}\:g^{4}\int\frac{d^{d}p}{(2\pi)^{d}}\frac{1}{(p^{2})^{3}}\int
\frac{d^{d}q}{(2\pi)^{d}}\frac{1+\cos p\:\theta q}{q^{2}(q+p)^{2}}
 \nonumber\\
&& =\frac{5}{2}\:g^{4}\big[\:{\cal F}(3,1)+{\cal
G}(3,1;\theta)\:\big]\eea

 \bea  \lambda_{4}\equiv
C_{6,0,4}&&\!\!\!\!\!\!\!\!=\frac{5}{2}\:g^{4}\int\frac{d^{d}p}{(2\pi)^{d}}\frac{1}{(p^{2})^{2}}\int
\frac{d^{d}q}{(2\pi)^{d}}\frac{1+\cos p\:\theta
q}{(q^{2})^{2}(q+p)^{2}}
 \nonumber\\
&& =\frac{5}{2}\:g^{4}\big[\:{\cal F}(2,2)+{\cal
G}(2,2;\theta)\:\big]\eea

In addition to terms with the coefficient $C_{n,0,i}(\Lambda_{NC})$,
i.e, the terms with nonderivative higher power of $\phi$, the higher
derivative terms could contribute to the effective Lagrangian. These
terms arise in a more exact treatment  when we stop neglecting the
external momenta of the diagrams. For instance, expanding $\phi$ in
(25), we could obtain the following terms:

 \bea
  \int d^{d}x \:\eta_{1}\phi(\partial\phi)^{3}\:\:,\:\: \int d^{d}x \:\eta_{2}\phi^{2}(\partial\phi)^{2}\:\:,\:\:
  \int d^{d}x \:\eta_{3}(\partial\phi)^{4}\:,\:\:.\:.\:. \eea

In general, the procedure of integrating out the fields $\hat{\phi}$
generate all possible interactions of the fields and their
derivatives. The coefficients $\eta_{1}$, $\eta_{2}$, $\eta_{3}$,
... are proportional to
 \beq \eta_{1},\: \eta_{2},\:
\eta_{3},\:...\propto\:\zeta_{1}+\:\zeta_{2}+\:\zeta_{3}  \eeq

Usually only nonderivative higher dimensional operators are
considered, while higher derivatives are less studied or less
popular due to the many difficult issues involved: either unitarity
or causality violation, non-locality and presence of ghost fields
with superluminal velocity \cite{14,22,23,24}. As already discussed
in \cite{14}, one can put a positivity constraint on higher order
irrelevant operators in order for the low energy effective theory to
be UV complete into a full theory correctly. At one loop level, we
see this positivity  satisfied for the non-derivative terms and we
have,
 \beq
C_{n,d,i}>0 \eeq

We must note that the positivity constraint is different from
other familiar positivity constrains that follow from vacuum
stability (for example, the positivity of the kinetic term and
also $m^{2}\phi^{2}$ and $g\phi^{4}$ couplings). The effective
models with irrelevant terms and a negative sign can not arise as
a low energy limit of any familiar UV-complete theories. The
positivity constraint at least at one loop level is satisfied for
the irrelevant derivative terms as is explicit from (38) and the
fact (48).

%%%%%%%%%%%%%%%%%%%%%%%%%%%%%%%%%%%%%%%%%%%%%%%%%%%%%%%%%%%%%%%%%%%%%%%%%%%%%%%%%%%%%%%%%%%%%%%%%%%%%%%%%%%%%%%%%%%%%%%%%%%
\subsection{$\phi^{3}$ theory}

We repeat the computations performed above for noncommutative
$\phi^{3}$ theory whose generating functional in d-dimensional
Euclidean space is
 \beq Z=\int {\cal D}\phi\exp
\Big(-\int d^{d}x
\Big[\:\frac{1}{2}\:\partial_{\mu}\phi\partial^{\mu}\phi+\frac{1}{2}m^{2}\phi^{2}
  +\frac{1}{3!}\:g\phi\star\phi\star\phi\:\Big]\Big)
 \eeq

It can be shown that in six dimensions the 1PI two point function
receives contribution from a one loop nonplanar diagrams for $
\tilde{p}\Lambda_{0}\ll 1$ and gets the one loop effective action as
\cite{6}
 \bea  S_{eff}=\int
d^{6}p\frac{1}{2}\:\Big(&&\!\!\!\!\!\!\!\!\!\!
p^{2}+M^{2}-\frac{g^{2}}{2^{\:8}\pi^{2}\tilde{p}^{2}
}\nonumber\\&&+\:\:\frac{g^{2}}{3\cdot
2^{\:9}\pi^{3}}(p^{2}+6M^{2})\ln(\frac{1}{M^{2}\tilde{p}^{2}
})+\cdot\cdot\cdot\Big)\phi(p)\phi(-p) \eea

 Where $M^{2}$ is the
mass renormalized by the planar diagram. The quadratic IR divergence
can be interpreted again in terms of new degrees of freedom $\chi$
as in \cite{6}. In spite of this, we try to rederive Wilsonian
effective action from our new formalism. So, as in $\phi^{4}$
theory, we divide the variables into commutative fields $\phi$ and
noncommutative fields $\hat{\phi}$. Then using the similar discussions of the
previous section, (55) gets the form,
 \bea
Z=\int {\cal D}\phi
e^{-\int\cal{L}(\phi)}&&\!\!\!\!\!\!\!\!\!\!\!\!\!\int{\cal
D}\h{\phi}\exp \Big(-\int d^{d}x
 \Big[\:\frac{1}{2}(\partial_{\mu}\h{\phi})^{2}+  \frac{1}{2}\:m^{2}\hat{\phi}\star\hat{\phi}\nonumber\\
&&
\:\:\:\:\:\:\:+g\:(\frac{1}{2}\:\phi\:\hat{\phi}\star\hat{\phi}+\frac{1}{2}\phi^{2}\hat{\phi}
      +\frac{1}{3!}\:\hat{\phi}\star\hat{\phi}\star\hat{\phi})\:\Big]\Big)
   \eea
Integrating out the high momentum noncommutative degrees of freedom
$\hat{\phi}$, we obtain the Wilsonian effective action
$S_{eff}(\phi;\Lambda_{NC})$ with the corresponding Lagrangian
density,
 \bea {\cal
L}_{eff}(\phi;\Lambda_{NC})\!\!\!\!\!\!\!\!\!&&=\frac{1}{2}Z(\Lambda_{NC})\partial_{\mu}\phi\partial^{\mu}\phi
+\frac{1}{2}\:m^{2}(\Lambda_{NC})\phi^{2}+\frac{1}{3!}\:g(\Lambda_{NC})\phi^{3}\nonumber\\&&
\:\:\:\:\:\:\:\:\:\:\:\:\:\:\:\:\:\:\:\:\:\:+\sum_{i}C_{n,d,i}(\Lambda_{NC}){\cal
O}_{n,d,i} \eea

 We can expand the exponential and use Wick's theorem,
to define the propagator and find Feynman rules. The fields
$\hat{\phi}$ contribute to internal lines and its propagator will be
the same as (28). The diagrams with two, three, and four external
lines are drawn in Figs. 4,5 and 6 respectively. Here we present the
expressions associated with the above diagrams in the following
form:

\beq Fig.\: 4=-\frac{1}{2}\int d^{d}x\:(\mu_{1}+\mu_{2}+\mu_{3})
\phi^{2} \eeq

\beq Fig.\:5=-\frac{1}{3!}\:\int
d^{d}x\:(\zeta_{1}+\zeta_{2}+\zeta_{3})\phi^{3} \eeq \beq
Fig.\:6=-\frac{1}{4!}\:\int
d^{d}x\:(\lambda_{1}+\lambda_{2}+\lambda_{3}+\lambda_{4})\phi^{4}.
\eeq

\begin{figure}

\centerline{\epsfysize=5in\epsfxsize=7in\epsffile{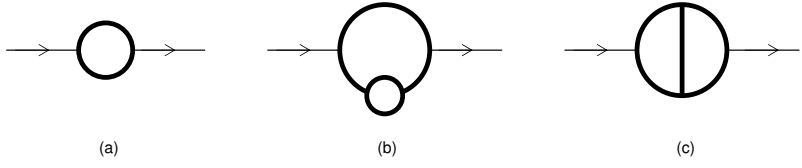}}\vspace{-7
cm }\caption{Two external line diagrams
 }\label{fig4}
\end{figure}

\begin{figure}

\centerline{\epsfysize=5.1in\epsfxsize=7.1in\epsffile{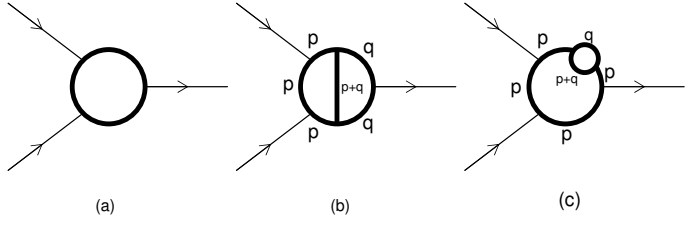}}\vspace{-7
cm }\caption{Three external line diagrams
 }\label{fig5}
\end{figure}

\begin{figure}

\centerline{\epsfysize=5in\epsfxsize=7in\epsffile{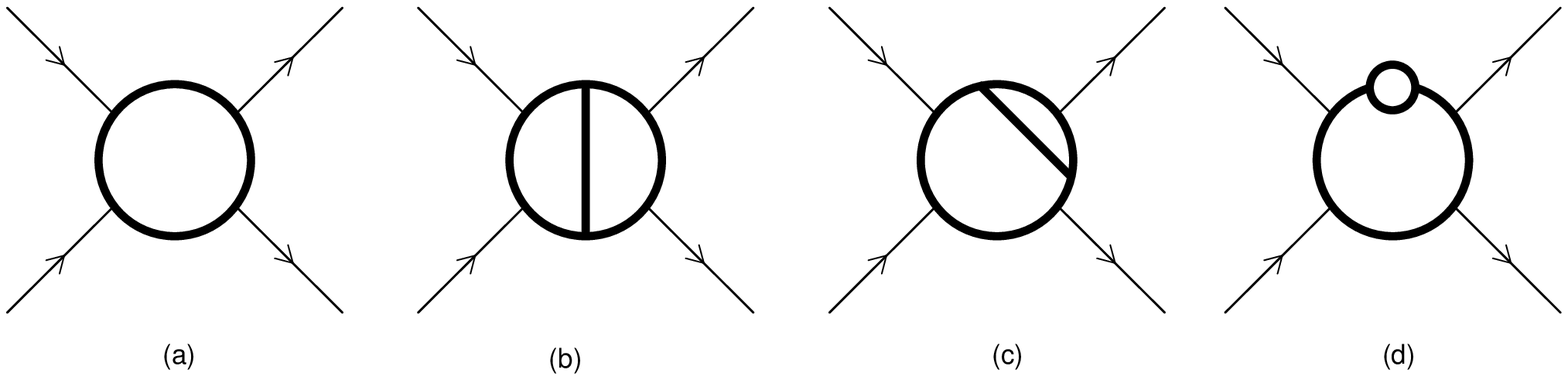}}\vspace{-7
cm }\caption{Four external line diagrams
 }\label{fig6}
\end{figure}

\noindent In Eq. (59), the terms $\mu_{i}$ are calculated
straightforwardly and the results are \beq
\mu_{1}=\frac{g^{2}}{2}\int\frac{d^{d}q}{(2\pi)^{d}}\frac{1}{(q^{2})^{2}}=g^{2}\frac{\pi^{\frac{d}{2}}}{(2\pi)^{d}\Gamma(\frac{d}{2})}\frac{1}{d-4}
  (\Lambda_{0}^{d-4}-\Lambda_{NC}^{d-4})
\eeq
 \bea \mu_{2}\!\!\!\!\!\!\!&&=\frac{g^{4}}{2\cdot
(3!)^{2}}\int\frac{d^{d}p}{(2\pi)^{d}}\frac{1}{(p^{2})^{3}}\int\frac{d^{d}q}{(2\pi)^{d}}\frac{1+\cos
p\theta q }{q^{2}(q+p)^{2}}\nonumber\\&&= \frac{g^{4}}{2\cdot
(3!)^{2} }\big[\: {\cal F}(3,1)+{\cal G}(3,1;\theta) \:\big]\eea
\bea \mu_{3}\!\!\!\!\!\!\!\!&&=\frac{g^{4}}{2\cdot
(3!)^{2}}\int\frac{d^{d}p}{(2\pi)^{d}}\frac{1}{(p^{2})^{2}}\int\frac{d^{d}q}{(2\pi)^{d}}\frac{1+\cos
p\theta q }{(q^{2})^{2}(q+p)^{2}} \nonumber\\&&=\frac{g^{4}}{2\cdot
(3!)^{2} }\:\big[\: {\cal F}(2,2)+{\cal G}(2,2;\theta) \:\big] \eea
These coefficients modify the $m^{2}$ term of ${\cal L}$ in (57) as,
\bea
m^{2}(\Lambda_{NC})=m^{2}(\Lambda_{0})\:+\:\mu_{1}+\mu_{2}+\mu_{3}\eea

\noindent Similarly, for the terms $\zeta_{1}$,$\zeta_{2}$ and
$\zeta_{3}$ we have,

 \beq
\zeta_{1}=3!\:\frac{g^{3}}{8}\int\frac{d^{d}q}{(2\pi)^{d}}
\frac{1}{(q^{2})^{3}}=\frac{3}{4}g^{3}\frac{2\pi^{\frac{d}{2}}}{(2\pi)^{d}\Gamma(\frac{d}{2})}\frac{1}{d-6}\big(\Lambda_{0}^{d-6}-\Lambda_{NC}^{d-6}\big)\eeq
\bea \zeta_{2}=\!\!\!\!\!\!\!\!&&=3\frac{g^{5}}{16\cdot
3!}\int\frac{d^{d}p}{(2\pi)^{d}}\frac{1}{(p^{2})^{3}}\int\frac{d^{d}q}{(2\pi)^{d}}\frac{1+\cos
p\theta q }{(q^{2})^{2}(q+p)^{2}}
\nonumber\\&&=3\frac{g^{5}}{16\cdot 3!}\:\big[\: {\cal F}(3,2)+{\cal
G}(3,2;\theta) \:\big] \eea \bea
\zeta_{3}=\!\!\!\!\!\!\!\!&&=3\frac{g^{5}}{16\cdot
3!}\int\frac{d^{d}p}{(2\pi)^{d}}\frac{1}{(p^{2})^{4}}\int\frac{d^{d}q}{(2\pi)^{d}}\frac{1+\cos
p\theta q }{q^{2}(q+p)^{2}} \nonumber\\&&=3\frac{g^{5}}{16\cdot
3!}\:\big[\: {\cal F}(4,1)+{\cal G}(4,1;\theta) \:\big] \eea
 The terms
$\zeta_{1}$, $\zeta_{2}$ and $\zeta_{3}$ give a correction to the
coupling g as, \bea
g(\Lambda_{NC})=g(\Lambda_{0})+\zeta_{1}+\zeta_{2}+\zeta_{3} \eea

Note how the effects of noncommutativity appear in the corrections
of $m^{2}$ and $g$. Here as in the $\phi^{4}$ theory, in $m^{2}$ and
$g$, the quantum correction is free of UV/IR mixing at the limit
$\theta\rightarrow 0$. The coefficients of higher order operators
(for example, $\phi^{4}$) can be derived by taking into account the
diagrams drawn in Fig. 6. For the coefficients $\lambda_{i}$, we
have \bea \lambda_{1}\equiv
C_{4,0,1}\!\!\!\!\!\!\!&&=\frac{g^{4}}{3!}\int\frac{d^{d}q}{(2\pi)^{d}}\frac{1}{(q^{2})^{\:4}}\nonumber\\&&
=\frac{g^{4}}{3!}\frac{2\pi^{\frac{d}{2}}}{(2\pi)^{d}\Gamma(\frac{d}{2})}
\frac{1}{d-8}\big(\Lambda_{0}^{d-8}-\Lambda_{NC}^{d-8} \big) \eea
\bea \lambda_{2}\equiv C_{4,0,2}\!\!\!\!\!\!\!&&=\frac{g^{6}}{2\cdot
3!}\int\frac{d^{d}p}{(2\pi)^{d}}\frac{1}{(p^{2})^{3}}\int\frac{d^{d}q}{(2\pi)^{d}}\frac{1+\cos
p\theta q }{(q^{2})^{3}(q+p)^{2}} \nonumber\\&&=\frac{g^{6}}{2\cdot
3!}\:\big[\: {\cal F}(3,3)+{\cal G}(3,3;\theta) \:\big]\eea \bea
\lambda_{3}\equiv C_{4,0,3}\!\!\!\!\!\!\!&&=\frac{g^{6}}{
3!}\int\frac{d^{d}p}{(2\pi)^{d}}\frac{1}{(p^{2})^{4}}\int\frac{d^{d}q}{(2\pi)^{d}}\frac{1+\cos
p\theta q }{(q^{2})^{2}(q+p)^{2}} \nonumber\\&&=\frac{g^{6}}{
3!}\:\big[\: {\cal F}(4,2)+{\cal G}(4,2;\theta) \:\big]\eea \bea
\lambda_{4}\equiv C_{4,0,4}\!\!\!\!\!\!\!&&=\frac{g^{6}}{
3!}\int\frac{d^{d}p}{(2\pi)^{d}}\frac{1}{(p^{2})^{5}}\int\frac{d^{d}q}{(2\pi)^{d}}\frac{1+\cos
p\theta q }{q^{2}(q+p)^{2}} \nonumber\\&&=\frac{g^{6}}{ 3!}\:\big[\:
{\cal F}(5,1)+{\cal G}(5,1;\theta) \:\big]\eea

In addition to the terms (59-61), we may have some higher
dimensional derivative terms. For instance, by expanding the fields
$\phi$ in (65) the expressions, \beq \int
d^{d}x\:\eta_{1}\phi(\partial\phi)^{2}\:\:,\:\:\int
d^{d}x\:\eta_{2}\phi^{2}\partial\phi\:\:,\:\: \int
d^{d}x\:\eta_{3}(\partial\phi)^{3}\:\:,\:\:\int
d^{d}x\:\eta_{4}\phi^{2}\partial^{2}\phi\:\:,...\:\:\eeq

may contribute to (57) with the coefficients $\eta_{i}$ proportional
to $\zeta_{1}+\zeta_{2}+\zeta_{2}$. The positivity constraint
\cite{14} on the coefficients of the derivative interaction terms
and also higher order operators are concluded from (64) and (70) up
to one loop level.

At the end of this section we briefly explain our method and compare
it with the theory of Grosse and Wulkenhaar \cite{11,12,a10,a11}. The
approach of the present work is based on the  fact that
noncommutativity is an effective picture of space-time in the Planck
scales where quantum gravity fluctuations change the classical
notion of space and time. So we expect that noncommutativity is
only relevant  at very high energies while in our ordinary scales,
space-time is  considered to be commutative. In \cite{11,12,a10,a11} the
term
$\frac{\Omega^{2}}{2}(\tilde{x}_{\mu}\phi)\star(\tilde{x}^{\mu}\phi)$
has been considered in order to get rid of UV/IR mixing problem.
This term has important effects only at large distances and is
irrelevant at short distances. So the low energy theory is affected
by noncommutativity but interactions can cure this problem and
finally leads the theory be free from UV/IR mixing problem. We have
shown that the same result can be obtained by assuming a commutative
space at low energies and by integrating out the high energy
noncommutative modes beyond the scale $\Lambda_{NC}$. In the high
energy  region existence of the harmonic oscillator term  does not
change the results of two previous subsection.

%%%%%%%%%%%%%%%%%%%%%%%%%%%%%%%%%%%%%%%%%%%%%%%%%%%%%%%%%%%%%%%%%%%%%%%%%%%%%%%%%%%%%%%%%%%%%%%%%%%%%%%%%%%%%%%%%%%%%%%%
\section{Noncommutative extra dimension}

We first discuss about extra dimension proposed in reference
\cite{25}. In a noncommutative theory in addition to noncommutative
quadratic IR singularities, there exist noncommutative logarithmic
singularities. The authors of \cite{25} have suggested some new
light degrees of freedom to interpret these logarithmic IR
singularities. For example, in $\phi^{3}$ theory the logarithmically
singular behavior of one loop effective action present in expression
(56) may be reproduced as a Wilsonian effective action arising from
the exchange of new scalar particles $\chi_{1}$ and $\chi_{2}$ which
couple to fields $\phi$ and have the propagators, \bea
&&\big<\chi_{1}(p)\chi_{2}(-p)\big>=-\frac{1}{3\cdot
2^{9}\pi^{3}}\ln\Big(\frac{\:\tilde{p}^{\:2}+\frac{1}{\Lambda^{2}\:}}{\tilde{p}^{\:2}}\:\Big)\nonumber\\&&
\big<\chi_{1}(p)\chi_{1}(-p)\big>=\big<\chi_{2}(p)\chi_{2}(-p)\big>=0
\eea

The $\chi_{i}$ fields may be replaced with the continuums of states
$\psi_{m}$ which have the ordinary propagators,
 \beq
\big<\psi_{m}(p)\psi_{m}(-p)\big>\propto \:
\frac{1}{\frac{\tilde{p}^{\:2 }}{\:\:(\alpha')^{2}}+m^{2}}\eeq

 \noindent where the
constant $\alpha'$ is defined as
$g^{\mu\nu}=-\frac{(\theta^{2})^{\mu\nu}}{\:\:(\alpha')^{2}}$. An
interpretation for these continuums of degrees of freedom $\psi_{m}$
is that they are the transverse momentum modes of a particle $\psi$
which propagates freely in extra dimensions. It may be imagined that
a d-dimensional space in which the $\phi$ particles propagate is a
flat d-dimensional brane residing in a d+n dimensional space. The
$\psi$ particles propagate freely in d+n dimensional space but
couple to the $\phi$ particles on the brane (located at
$x_{\perp}=0$). Actually the theory with free $\psi$ particles in
extra dimensions and coupling directly to $\phi$\:s on the brane is
exactly equivalent to a theory with particles $\chi$ that live on
the brane and have logarithmic propagators \cite{25}. These extra
dimensions may be real or simply a mathematical convenience. In
spite of this, it is not necessary in our approach to introduce
extra dimensions since the Wilsonian effective action that we
rederived is free from noncommutative IR singularities. Of course,
even without continuum states $\psi_{m}$, the world may have one or
more compactified extra dimensions. It is possible that such compact
dimensions are noncommutating \cite{26,27}. In the rest of this
section, we will discuss these noncommutative extra dimensions in
$\phi^{3}$ theory on $R^{1,3}\times T_{\theta}^{2}$ (\cite{26})
where $T_{\theta}^{2}$ is a noncommutative two-torus whose
coordinates $x^{4}$ and $x^{5}$ satisfy \beq
\big[x^{4},x^{5}\big]=i\theta. \eeq

Since coordinates $x^{4}$ and $x^{5}$ have been compactified on a
rectangular torus with boundary conditions $0\leq x^{4},x^{5}\leq
2\pi R$, the momentum along the compact dimensions are quantized as
${\bf p}=\frac{{\bf n}}{R}$ where ${\bf n}=(n_{4},n_{5})$ are
integers. To perform the perturbative calculation, it is better to
separate field $\phi$ as the commutative fields $\phi_{0}$
containing ${\bf n}=0$ and the noncommutative fields
$\overline{\phi}$ containing all modes with nonvanishing ${\bf n}$.
We make this separation as follows,
 \beq
 \phi_{0}=\frac{1}{(2\pi R)^{2}}\int dx^{4}dx^{5}\:\phi\;, \eeq
and
 \beq \overline{\phi}=\phi-\phi_{0}. \eeq
 In terms of these definitions, the partition function (57) in six
 dimensions gets the form,
 \bea Z=\int {\cal D}\phi \exp\big(-\int d^{6}x \big[ \frac{1}{2}(\partial\overline{\phi})^{2}- &&\!\!\!\!\!\!\!\!\! \frac{1}{2}m^{2}\overline{\phi}^{2}
        +\frac{1}{2}(\partial\phi_{0})^{2}-\frac{1}{2}m^{2}\phi_{0}^{2}
           -\frac{g}{3!}\overline{\phi}\star\overline{\phi}\star\overline{\phi}\nonumber\\&&
           -\frac{g}{2}\phi_{0}\overline{\phi}\star\overline{\phi}
           -\frac{g}{3!}\phi_{0}^{3} \:\big] \big).
 \eea

Note that the relation between compactification scale
$\Lambda_{c}=\frac{1}{R}$ and $\Lambda_{NC}$ is unknown. Naturally,
there are two possibilities about their relation:
$\Lambda_{c}<\Lambda_{NC}$ or $\Lambda_{c}\geq\Lambda_{NC}$. If
$\Lambda_{c}<\Lambda_{NC}$ (for example $\frac{1}{R}<$ few TeV),
 the extra dimensions are first probed and then noncommutative effects appear continuously beyond the scale
$\Lambda_{NC}$. So, as in the previous section, we can separate the
$\overline{\phi}$ fields into commutative $\overline{\phi}$ fields
and noncommutative $\overline{\phi}_{NC}$ fields with the momentum
$\Lambda_{NC}<k<\Lambda_{0}$. From the Wilsonian point of view, the
$\overline{\phi}_{NC}$ fields are integrated out when we calculate
the effective action. So we get an ordinary $\phi^{3}$ theory
compactified on $T^{2}$. The noncommutativity parameter $\theta$
appears in renormalized mass and coupling constant with a series of
higher order irrelevant operators. On the other hand, if
$\Lambda_{c}\geq \Lambda_{NC}$, the situation will be different.
Here, the extra dimensions are noncommutative at all scales. In this
region, the perturbation calculation of nonplanar diagrams shows
that the one loop self energy for $\bf{n}\neq 0$ gets the form \cite{26},
 \beq \Sigma=
-\frac{g^{2}}{(4\pi)^{3}}\:\big(\:\frac{R^{2}}{\theta^{2}{\bf
n}^{2}}\:+\:\frac{5}{24}m^{2}\ln\big(\frac{m^{2}\theta^{2}{\bf
n}^{2}}{R^{2}}\big)+\cdot\cdot\cdot \big) \eeq

So the one loop effective action $S_{eff}$ contains quadratic and
logarithmic IR singularities. While if $R$ is so large that
$\Lambda_{c}< \Lambda_{NC}$, the effective action is free from the
UV/IR mixing problem.

%%%%%%%%%%%%%%%%%%%%%%%%%%%%%%%%%%%%%%%%%%%%%%%%%%%%%%%%%%%%%%%%%%%%%%%%%%%%%%%%%%%%%%%%%%%%%%%%%%%%%%%%%%%%%%%%%%%%%%%%%%%
\section{Conclusion}

We have derived a consistent Wilsonian effective field theory for
noncommutative $\phi^{4}$ and $\phi^{3}$ theories. The effects of
noncommutativity appears in mass, coupling constant and the
$C_{n,d,i}$, the  coefficients of higher order operators. We have
shown that the noncommutative effective $\phi^{3}$ and $\phi^{4}$
theories are free from the  UV/IR mixing problem in the limit
$\theta\rightarrow 0$. Similarly, all order calculations  can also
have a soft behaviour and this result is natural in consequence of
our assumption that beyond the scale $\Lambda_{NC}$, the
noncommutative effects can be probed. Below this scale, space-time
becomes commutative continuously. We also studied the positivity
constraint on the coefficients of the higher order irrelevant
operator. These coefficients satisfy this constraint up to one
loop explicitly. The noncommutative extra dimensions were
considered in the noncommutative $\phi^{3}$ model. Our results
show that if the extra dimensions are large such that
$\Lambda_{c}=\frac{1}{R}<\:\Lambda_{NC}$, the noncommutative IR
divergences or UV/IR mixing could  disappear. For a similar idea
to what is proposed in this paper but in a different context see
\cite{19}, and also \cite{28}.

\section{Acknowledgment}

We would like to thank M.M. Sheikh-Jabbari for  useful discussion
and suggestions.

\section{Appendix \:\:A }

The functions ${\cal F}(n,m)$ and ${\cal G}(n,m;\theta)$ are defined
as follows \bea {\cal F}(n,m)
\!\!\!\!\!\!\!&&=\int\frac{d^{d}p}{(2\pi)^{d}}\frac{1}{(p^{2})^{n}}\int\frac{d^{d}q}{(2\pi)^{d}}\frac{1}{(q^{2})^{m}(q+p)^{2}}
\nonumber\\&& =
\frac{1}{2}\Big(\frac{2\:\pi^{\frac{d}{2}}}{(2\pi)^{d}\Gamma(\frac{d}{2})}\Big)^{2}\sum_{r=0}^{\infty}(-1)^{r}\frac{\Gamma(m+r+1)}{\Gamma(m)r!(r+\frac{d}{2})}
\int_{0}^{1}dx\:\frac{1}{x^{m}(1-x)^{r+2}}\nonumber\\&&
\:\:\:\:\:\times\int_{\Lambda_{NC}}^{\Lambda_{0}}
dp\:p^{d-2(n+m+r-1)-1}
\big[\:(\Lambda_{0}+xp)^{r+\frac{d}{2}}-(\Lambda_{NC}+xp)^{r+\frac{d}{2}})\:\big]\eea
 and
 \bea {\cal G}(n,m;\theta)=\!\!\!\!\!\!\! &&\int\frac{d^{d}p}{(2\pi)^{d}}\frac{1}{(p^{2})^{n}}\int\frac{d^{d}q}{(2\pi)^{d}}\frac{\cos
p\theta q }{(q^{2})^{m}(q+p)^{2}} \nonumber\\&&=
\frac{1}{2}\Big(\frac{2\:\pi^{\frac{d}{2}}}{(2\pi)^{d}\Gamma(\frac{d}{2})}\Big)^{2}
\frac{1}{\Gamma(m)}
\sum_{r=0}^{\infty}(-1)^{r}\frac{1}{r!(r+\frac{d}{2})}
\int_{0}^{1}(1-x)^{m}\int_{0}^{\infty}d\tau\:\tau^{m+r}\nonumber\\&&\times
\int_{\Lambda_{NC}}^{\Lambda_{0}}dp\:p^{d-2n-1}
\exp\big(-\tau\:x(1-x)p^{2}-\frac{1}{4\tau}\tilde{p}^{2}\big)
\nonumber\\&&
\times\Big[(\Lambda_{0}+\frac{i}{2\tau}\tilde{p}+xp)^{r+\frac{d}{2}}-(\Lambda_{NC}+\frac{i}{2\tau}\tilde{p}+xp)^{r+\frac{d}{2}}
 \Big]\eea

%%%%%%%%%%%%%%%%%%%%%%%%%%%%%%%%%%%%%%%%%%%%%%%%%%%%%%%%%%%%%%%%%%%%%%%%%%%%%%%%%%

\end{document}